\documentclass[a4paper]{article}
\usepackage{RR}
\RRNo{6312}
\usepackage{hyperref}
\usepackage[latin1]{inputenc}
 \usepackage{amssymb,amsbsy,amsmath,float,epsfig,here}
  \usepackage{subfigure,epsfig}
  \usepackage{amssymb,amsmath}
\RRdate{Octobre 2007} 

\RRauthor{
Taous-Meriem Laleg 
 \and
 Claire Médigue 
\and
Yves Papelier 
\and
Emmanuelle Crépeau 
  \and
Michel Sorine 
} 
\authorhead{Laleg \& Médigue \& Papelier \& Crépeau  \& Sorine}
\RRtitle{Nouveaux indices cardio-vasculaires basés sur une analyse spectrale
non linéaire de la forme des ondes de pression artérielle\\
}
\RRetitle{New Cardiovascular Indices Based on a Nonlinear Spectral
Analysis of Arterial Blood Pressure Waveforms} 
\titlehead{\small{New cardiovascular indices based on a nonlinear spectral analysis of arterial blood pressure waveforms}}
\RRresume{Une nouvelle méthode d'analyse du signal de pression
artérielle est proposée dans ce rapport. La technique est basée
sur la transformation scattering et consiste à résoudre un
problème spectral associé à l'opérateur de Schrodinger
unidimensionnel avec un potentiel qui dépend linéairement de la
pression. Le potentiel est reconstruit à l'aide du spectre discret
uniquement qui comprend des valeurs propres négatives
correspondant à N solitons en interaction. L'approche est analogue
à une transformée de Fourier où les solitons jouent le rôle des
composantes sinus et cosinus. La méthode introduit de nouveaux
indices cardio-vasculaires qui semblent contenir des informations
physiologiques intéressantes. Dans un premier temps, nous
appliquons la méthode à la décomposition du signal de pression
artérielle en ondes élémentaires et à sa reconstruction ou à la
séparation de ses parties diastolique et systolique. Nous
analysons par la suite les paramètres calculés à partir de cette
technique dans deux conditions physiologiques généralement
utilisées pour l'étude du contrôle cardio-vasculaire à
court-terme: test d'inclinaison à 60 degrés et exercice
isométrique (handgrip). Des résultats prometteurs ont été obtenus.
}

\RRabstract{A new method for analyzing arterial blood pressure is
presented in this report. The technique is based on the scattering
transform and consists in solving the spectral problem associated
to a one-dimensional Schrödinger operator with a potential
depending linearly upon the pressure. This potential is then
expressed with the discrete spectrum which includes negative
eigenvalues and corresponds to the interacting components of an
N-soliton. The approach is similar to a nonlinear Fourier
transform where the solitons play the role of sine and cosine
components. The method provides new cardiovascular indices that
seem to contain relevant physiological information. We first show
how to use this approach to decompose the arterial blood pressure
pulse into elementary waves and to reconstruct it or to separate
its systolic and diastolic phases. Then we analyse the parameters
computed from this technique in two physiological conditions, the
head-up 60 degrees tilt test and the isometric handgrip test,
widely used for studying short term cardiovascular control.
Promising results are obtained.} 

\RRmotcle{Pression artérielle, transformation scattering,
solitons, indices cardio-vasculaires} \RRkeyword{Blood pressure,
scattering transform, solitons, cardiovascular indices} 
 \RRprojet{SISYPHE}  
 \RRtheme{\THBio} 
 \URRocq 
\begin{document}
\makeRR   

\tableofcontents
\newpage
\section{Introduction}

The analysis of mean values and beat-to-beat variability of
cardiovascular (CV) time series has been widely used as a
noninvasive approach to study the control of the autonomic nervous
system (ANS) on the CV function \cite{BaBrLo:06},
\cite{MaMoMe:01}. CV time series usually give information in the
frequency and amplitude domains. The frequency (or period) is
given by the RR interval (between two R peaks on the ECG ) or the
pulse interval (PI) (between two systolic blood pressure peaks).
The amplitude concerns systolic, diastolic and mean pressures
(noted SBP, DBP and MBP respectively). Standard measures of these
parameters are mean levels and global variability, spectral,
temporal and time-frequency analysis \cite{MoMeMa:02}.\\

Instead of the usual decomposition of the arterial blood pressure
(ABP) waveform into a linear superposition of harmonic waves (sine
and cosine functions) \cite{SeVe:02}, \cite{WeSiVaEl:72}, in this
article we propose to use nonlinear superpositions of particular
travelling waves, the N-solitons-solutions of the Korteweg-de
Vries (KdV) equation where N travelling components are
interacting. These N-solitons play the role here of the
harmonic-waves solutions of the linear wave equation. The concept
of soliton refers in fact to a solitary wave emerging unchanged in
shape and speed from the collision with other solitary waves
\cite{Re:99}. They fascinate scientists by their very interesting
coherent-structure characteristics and are used in many fields to
model natural phenomena. Solitons are solutions of nonlinear
dispersive equations like the KdV equation arising in a variety of
physical problems, for example to describe wave motion in shallow
water canals \cite{ScChMc:73}, \cite{ZaKr:65}. The use of solitons
to describe the ABP was already introduced in \cite{Yom:87} and
\cite{PaRe:94} where a KdV equation and a Boussinesq equation were
respectively proposed as a blood flow model. Recently, in
\cite{CrSo:07}, \cite{LaCrSo:07J} a reduced model of the ABP cycle
was introduced. The latter consists of a sum of a 2 or 3-soliton
solution of a KdV equation, describing fast phenomena during the
systolic phase and a 2-element windkessel model describing slow
phenomena during the diastolic phase. We recall that the systolic
phase corresponds to the contraction of the heart, driving blood
out of the left ventricle while the diastolic phase corresponds to
the  period of relaxation of the heart.\\

The decomposition of the ABP signal into a nonlinear superposition
of solitons introduced in this article is based on an elegant
mathematical transform: the scattering transform for a
one-dimensional Schrödinger equation \cite{CaDe:82},
\cite{EcVa:83}, \cite{GaGrKrMi:74}. The main idea in our
utilization of this transform consists in interpreting the
pressure as a potential able to attract or repulse "fluid
particles" or equivalently to transmit or reflect waves associated
with them. This situation is modelled by a one-dimensional
Schrödinger operator with a potential depending linearly upon the
pressure wave \cite{LaCrSo:07}.  The discrete levels of energy or
speed of this system are given by the discrete spectrum of the
Schrödinger operator. The associated eigenstates describe some
coherent structures that are present in the pressure waveform and
expressed as N solitons.\\

The scattering-based signal analysis (SBSA) method introduces a
new spectral description of the pressure waveform leading to new
cardiovascular indices. This study aims to analyse these new
parameters in two physiological conditions that are widely used
for studying the short term control of the CV system: the head-up
60 degrees tilt test, in 15 healthy subjects, and the isometric
handgrip exercise, in 13 healthy subjects.\\

Among the SBSA  parameters,  some invariants of the scattering
transform seem to be related to the contraction strength of the
left ventricle and the stroke volume (SV). Moreover, the
beat-to-beat relation between the first eigenvalue and the heart
period might be more
reliable than the baroreflex slope for distinguishing two conditions.\\

In the next section, we present the basis of the SBSA method.
Section III compares real and reconstructed pressures using the
SBSA and shows how we can separate the fast and slow pressure
components related to the systolic and diastolic phases
respectively. Section IV introduces the new cardiovascular indices
computed using the SBSA technique and presents the results of the
analysis in two physiological conditions: tilt and handgrip,
including a discussion. Finally a conclusion summarizes the
different results.

\section{A scattering-based signal analysis method}
In this section, we introduce a new signal analysis method based
on the scattering theory. We start by briefly recalling the basis
of the Direct and Inverse Scattering Transforms (DST \& IST).
Then, we present the main idea in the SBSA technique. For more
details about DST and IST the reader can refer to the abundant
literature and the references given.


\subsection{Scattering transform for a Schrödinger equation}
Let $V$ be a given real function in the so-called Faddeev class
$L^1_1(\mathbb{R})$ \cite{AkKl:01}:
\begin{equation}\label{condpotentiel}
L_1^1(\mathbb{R})= \{V\in
L^1(\mathbb{R}),\int_{-\infty}^{+\infty}{|V(x)|(1+|x|)
dx}<\infty\}.
\end{equation}
We consider the one dimensional Schrödinger operator $H(V)$ with a
potential $V$:
\begin{equation}
H(V): \psi \rightarrow H(V)\psi = -\frac{\partial^2 \psi}{\partial
x^2} + V \psi.
\end{equation}
The DST of $V$ will be defined as a function of the solution of
the spectral problem for $H(V)$ where $\lambda$ and $\psi$ are
respectively the eigenvalues and the associate eigenfunctions for
some normalization:
\begin{equation}\label{shr1}
   H(V)\psi = \lambda  \psi. 
\end{equation}
The spectrum of $H(V)$ has two components: a continuous spectrum
equal to $\left( 0, +\infty \right)$ and a discrete spectrum with
negative eigenvalues \cite{DeTr:79}, \cite{EcVa:83},
\cite{GaGrKrMi:74}.\\

For the positive eigenvalues denoted here $\lambda=k^2$, there are
eigenfunctions, called scattering solutions of (\ref{shr1})
consisting of linear combinations of $\exp{(ikx)}$ and
$\exp{(-ikx)}$ as $x\rightarrow \pm\infty$. Among them, we
consider the Jost solutions from the left $f_l$ and from the right
$f_r$, normalized at $\pm \infty$:
\begin{eqnarray}\label{equ3bis}
    H(V)f_j \:\:\:\:\: \: & = & k^2f_j, \:\:\:\: k\in \mathbb{R}\backslash\{0\},
    \:\:\:\: j=l,r,\\
  \exp{(-ikx)}f_l(k,x)\:\:\:\:\: \:& = &1+o(1),\quad \:\:\:\:\: \: \:\:x\rightarrow +\infty,\\
  \exp{(-ikx)}\frac{\partial f_l(k,x)}{\partial x} &= &ik+o(1), \:\:\:\:\: \:
  \quad x\rightarrow +\infty,\\
  \exp{(+ikx)}f_r(k,x) \:\:\:\:\: \:&= &1+o(1), \quad \:\:\:\:\: \: \:\: x\rightarrow -\infty,\\
  \exp{(+ikx)}\frac{\partial f_r(k,x)}{\partial x}& =& -ik+o(1), \quad x\rightarrow -\infty,
\end{eqnarray}
We recall that, for each fixed $x\in \mathbb{R}$ the Jost
solutions have analytic extensions in $k$ to the upper-half
complex plane \cite{AkKl:01}.\\

The transmission coefficient $T$ and the reflection coefficients
$R_l$ and $R_r$ from the left and from the right respectively are
defined through the relations:
\begin{equation}
  f_l(k,x)=\frac{\exp{(ikx)}}{T(k)}+\frac{R_l(k)\exp{(-ikx)}}{T(k)}+o(1), \:\: x\rightarrow -\infty,
  \end{equation}
  \begin{equation}
  f_r(k,x) = \frac{\exp{(-ikx)}}{T(k)}+\frac{R_r(k)\exp{(ikx)}}{T(k)}+o(1),
  \:\:\:   x\rightarrow +\infty,
\end{equation}
These coefficients satisfy:
\begin{equation}
    |T(k)|^2+|R_l(k)|^2=|T(k)|^2+|R_r(k)|^2=1.
\end{equation}

On the other hand, for the negative eigenvalues of the discrete
spectrum, equation (\ref{shr1}) admits solutions called bound
states that belong to $L^2(\mathbb{R})$ in the $x$ variable. When
$V$ belongs to the Faddeev class, the bound states solutions of
(\ref{shr1}) decay exponentially as $x\rightarrow \pm \infty$ and
their number $N$ is finite \cite{AkKl:01}. Let us denote
$\lambda_n = -\kappa_n^2$ with $\lambda_1 \leq \lambda_2 \leq ...$
and $\psi_n$ the $N$ negative eigenvalues and $L^2$-normalized
bound states:
\begin{equation}
    H(V)\psi_n=-\kappa_n^2\psi_n,\quad
    \int_{-\infty}^{+\infty}{|\psi_n(x)|^2dx}=1,\quad
    n=1,\cdots,N.
\end{equation}
The eigenspaces being of dimension $1$, the bound states and the
Jost solutions are proportional:
\begin{equation}
  \psi_n(x)=c_{ln}f_l(i\kappa_n,x)=(-1)^{N-n}c_{rn}f_r(i\kappa_n,x),
\end{equation}
where $c_{ln}$ and $c_{rn}$ are called the bound-state norming
constants and are defined by:
\begin{eqnarray}
    c_{jn}:=[\int_{-\infty}^{+\infty}{|f_j(i\kappa_n,x)|^2
    dx}]^{-\frac{1}{2}}, \quad j=l,r.
\end{eqnarray}
We now define the DST of $V$ as the sets of scattering data from
the left, $\mathcal{S}_l(V)$ or from the right,
$\mathcal{S}_r(V)$:
\begin{eqnarray}
\mathcal{S}_j(V) : = \{R_j, \; \kappa_n, \; c_{\bar{j}n}, \;\;
 n=1,\cdots,N \}, \quad \{j,\bar{j}\}=\{l,r\}.
\end{eqnarray}
The potential can be uniquely reconstructed by using any one of
these sets. The solution of this inverse problem, called IST, is
the object of many studies concerned with specific classes of
potentials \cite{AkKl:01}, \cite{ChSa:89}, \cite{DeTr:79},
\cite{GeLe:55}, \cite{Mar:86}. Two transforms are then available,
$\mathcal{S}$ and $\mathcal{S}^{-1}$ (in the sequel we choose
$j=r$ and drop the subscripts $r$ and $l$ for simplicity).\\

In this study, we will use the special class of {\em
reflectionless potentials} for which the left or right reflection
coefficients are zero. Such potentials can be constructed as
follows: let $\Pi_d$ be the projector zeroing the $R$-component of
$\mathcal{S}(V)$, then $\mathcal{S}^{-1} \circ \Pi_d \circ
\mathcal{S}(V)$ is reflectionless for any $V$ in the Faddeev
class. There are useful explicit representations of reflectionless
potentials using only the discrete spectrum as in the following
theorem \cite{GaGrKrMi:74}:\\

\textbf{\emph{Theorem}:} If $V$ is reflectionless for $H(V)$,
then:
\begin{equation}\label{pot5}
V(x)=-4\sum_{n=1}^{N}{\kappa_n\psi_n^2(x)}, \quad x\in \mathbb{R}.
\end{equation}
$V$ can be also written:
\begin{equation}\label{pot2}
V(x)=-2\frac{\partial^2(\log{(\det(I+A))})}{\partial x^2},\quad
x\in \mathbb{R},
\end{equation}
where $A$ is an $N \times N$ matrix:
\begin{equation}\label{pot22}
A(x)=[\frac{c_{m}
c_{n}}{\kappa_m+\kappa_n}\exp{((\kappa_m+\kappa_n)x)}], \quad
n,m=1,\cdots,N.
\end{equation}
Note that in (\ref{pot2}) and (\ref{pot22}), the potential is
entirely defined with $2N$ parameters namely $\kappa_n$ and
$c_{n}$, $n=1,\cdots,N$.\\

A very close relation between a soliton solution of a KdV equation
and a reflectionless potentiel of the Schrödinger operator was
introduced in \cite{GaGrKrMi:74}. In fact these potentials remain
reflectionless when evolving in time and space according to a KdV
equation. For $t \rightarrow +\infty$, $N$ 1-solitons emerge, each
one being characterized by a pair $(\kappa_n, c_n)$ such that
$4\kappa_n^2$ gives the speed of the soliton and $c_{n}$ its
position. Therefore each component $-4\kappa_n\psi_n^2$ in the sum
(\ref{pot5}) refers to a single soliton.

\subsection{A scattering based signal analysis method}
We now present how to use IST in a seemingly new
method to analyse pulse-like signals of the Faddeev class.\\

The main idea in the SBSA approach is to interpret a positive
signal $y$ in the Faddeev class as a quantum well by changing the
sign, and to tune the depth of this well with a positive parameter
$\chi$ in order to approximate $y$ by a coherent state $y_{\chi}$.
For a deeper well the trapped energy will be higher and the
approximation better, as we will prove.  The estimate is then
obtained by filtering out the nonlinear reflections:
\begin{equation}\label{projd}
y_{\chi} = -\frac{1}{\chi} \mathcal{S}^{-1} \circ \Pi_d \circ
\mathcal{S}(-\chi y).
\end{equation}
A convenient explicit formula is available, $\chi y_{\chi}$ being
a reflectionless potential:
\begin{equation}\label{y_chi}
   y_\chi = \frac{4}{\chi}\sum_{n=1}^{N_\chi}{\kappa_{\chi,n}\psi_{\chi,n}^2},
\end{equation}
where $-\kappa_{\chi,n}^2$ and $\psi_{\chi,n}$,
$n=1,\cdots,N_\chi$ are the negative eigenvalues and the
associated $L^2$-normalized eigenfunctions for $H(-\chi y)$.\\

Then, we look for a value $\hat{\chi}$ for the parameter $\chi$
such that the signal $y$ is well approximated by $y_{\hat{\chi}}$.
This is the decomposition of the signal $y$ into the nonlinear
superposition of solitons announced  in \cite{LaCrSo:07}.\\

\begin{figure}[tb]
  \begin{center}
  \includegraphics[width=8cm]{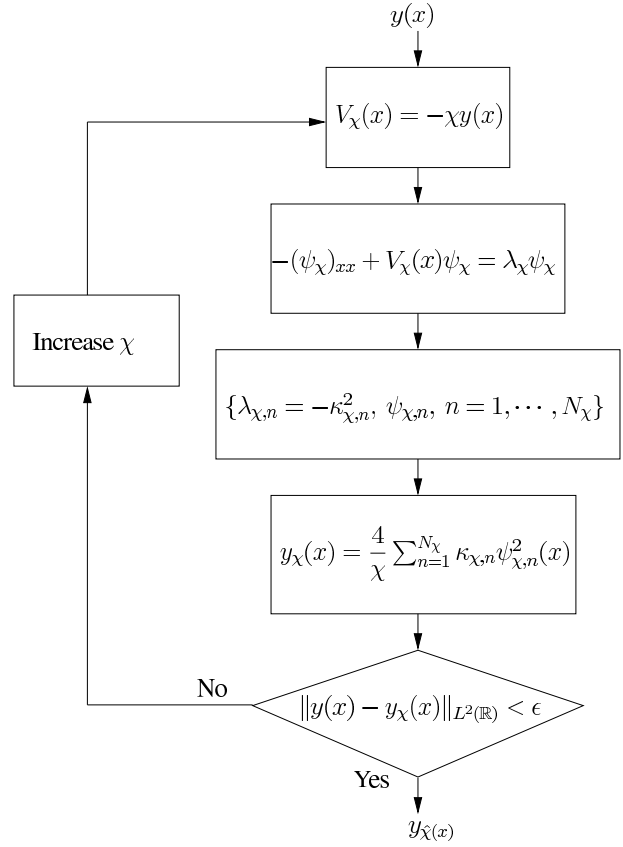}\\
  \caption{Signals analysis with the SBSA method}\label{principe}\end{center}
\end{figure}

It is well-known that the number of negative eigenvalues $N_\chi$
of $H(-\chi y)$  is a nondecreasing function of $\chi$ and there
is an infinite unbounded sequence $(\chi_n)$ such that
$N_{\chi_n}=N_{\chi_{n-1}}+1$ \cite{LaCrSo:07}, \cite{MiCl:01}.
Determining the parameter $\hat{\chi}$ determines the number of
negative eigenvalues and hence the number of solitons components
required for a satisfying approximation of the signal $y$. Fig.
\ref{principe} summarizes the SBSA technique.
%
\subsection{SBSA and the invariants of KdV}
The scattering transform has an infinite number of invariants
which are related to the KdV conserved quantities
\cite{GaGrKrMi:74}, \cite{MoNoVa:01}. Let us denote these
invariants $ I_m(V)$, $m=0,1,2,\cdots$. They are of the form (we
take $-V$ as argument having (\ref{projd}) in mind):
\begin{equation}\label{gen1}
I_m(-V)=(-1)^{m+1}\frac{2m+1}{2^{2m+2}}\!\!\int_{-\infty}^{+\infty}{P_m(V,\frac{\partial
V}{\partial x},\frac{\partial^2 V}{\partial x^2}, \cdots)dx},
\end{equation}
where $P_m$, $m=0,1,2,\cdots$ are known polynomials in $V$ and its
successive derivatives with respect to $x\in\mathbb{R}$
\cite{CaDe:82}.\\

A general formula relating $I_m(-V_\chi)$, with $V_\chi=-\chi y$,
to the scattering data of $H(V_\chi)$ can be deduced; see for
example \cite{CaDe:82}, \cite{GeHo:94}, \cite{MoNoVa:01}:
\begin{eqnarray}\label{invgeneral}
I_m(\chi y)\!\!=\!\!\sum_{n=1}^{N_\chi}
    {{\kappa_{\chi,n}^{2m+1}}}\!\!+\!\!\frac{2m+\!\!1}{2\pi}\!\!\!\int_{-\infty}^{+\infty}{\!\!\!(-k^2)^m \ln{(|T_{\chi}(k)|)}dk},
\end{eqnarray}
$ m=0,1,2,\cdots$.\\

We introduce the Riesz means of the negative eigenvalues
$\lambda_n$ of $H(V)$ such that $\lambda_n \leq \lambda \leq 0$:
\begin{equation}
S_{\gamma,\lambda}(-V)=\sum_{\lambda_n\leq
\lambda}{|\lambda_n|^\gamma}, \quad \gamma \geq 0.
\end{equation}
Remark that $S_{0,\lambda}(V)$ is the number of eigenvalues of
$H(V)$ smaller than $\lambda$.\\

For an $N_\chi$-soliton, for instance $-\chi y_\chi$ of the
previous subsection, the invariants only depend on the discrete
spectrum and they are related to the Riesz means as follows:
\begin{equation}
I_m(\chi y_\chi)=S_{\gamma,0}(\chi y_\chi), \quad
\gamma=m+\frac{1}{2},\quad m=0,1,2,\cdots
\end{equation}
A "sum rule" is then verified by the invariants of $\chi y$ and
$\chi y_{\chi}$:
\begin{equation}\label{gen2} I_m(\chi y)= I_m(\chi
y_\chi)\!\!+\!\!
\frac{2m+1}{2\pi}\!\!\!\int_{-\infty}^{+\infty}{\!\!\!(-k^2)^m
\ln{(|T_{\chi}(k)|)}dk},
\end{equation}
$m=0,1,2,\cdots$.\\

In this article we are only interested in the two first invariants
($m=0$ and $m=1$) corresponding to the conservation of mass and
momentum for the KdV flows. Here it is sufficient to see them as
invariants of the DST, in the same manner energy is invariant for
the Fourier transform (Plancherel's theorem). We will show later
in the application of the SBSA to the ABP that these two
invariants are related to some important cardiovascular
parameters.\\

So, for $m=0$, $P_0(V_\chi,\cdots)=V_\chi$, we get with
(\ref{gen1}) and (\ref{gen2}):
\begin{equation}
\int_{-\infty}^{+\infty}{y dx}=\int_{-\infty}^{+\infty}{y_{\chi}
dx} +  \frac{2}{\pi
    \chi} \int_{-\infty}^{+\infty}{\!\!\!\ln{(|T_{\chi}(k)|)}
    dk}.
\end{equation}
For $m=1$, $P_1(V_\chi,\cdots)=V_\chi^2$,  we have with
(\ref{gen1}) and (\ref{gen2}):
\begin{equation}\label{BFZ}
\int_{-\infty}^{+\infty}{y^2
dx}=\int_{-\infty}^{+\infty}{y_{\chi}^2 dx} - \frac{8}{\pi
    \chi^2} \int_{-\infty}^{+\infty}{\!\!\!k^2 \ln{(|T_{\chi}(k)|)}
    dk}.
\end{equation}
Equation (\ref{BFZ}) is known as the Buslaev-Faddeed-Zakharov
trace formula.\\

\textbf{\emph{Proposition}:} Let $y : \mathbb{R} \rightarrow
\mathbb{R}$ be a continuous non-negative function with a compact
support, then we have the convergence of the estimates of the
first two invariants:
\begin{equation}
\lim_{\chi\rightarrow +\infty}{I_m(y_\chi)= I_m(y)}, \quad m=0,1.
\end{equation}
Proof: We can apply the results on the Lieb-Thirring semiclassical
limit of the Riesz means \cite{BlSt:96},
 \cite{HeRo:90b}, \cite{HeRo:90a}, \cite{LaWe:00}:
\begin{equation}
    \lim_{\chi \rightarrow +\infty}{\frac{S_{\gamma,0}(\chi
    y)}{\chi^{\frac{1}{2}+\gamma}}}= L_{\gamma}^{cl}
    \int_{\mathbb{R}}{y(x)^{\frac{1}{2}+\gamma}dx}, \quad
    \gamma\geq \frac{1}{2},
\end{equation}
where $ L_{\gamma}^{cl}$ is the so-called Lieb-Thirring constant
given by:
\begin{equation}
L_{\gamma}^{cl}\equiv(4\pi)^{-\frac{1}{2}}\dfrac{\Gamma(\gamma+1)}{\Gamma(\gamma+\dfrac{3}{2})}.
\end{equation}
We notice that for $\gamma=\dfrac{1}{2}$:
\begin{equation}
    \lim_{\chi \rightarrow +\infty}{\frac{S_{\frac{1}{2},0}(\chi
    y)}{\chi}}= L_{{\frac{1}{2}}}^{cl}
    \int_{\mathbb{R}}{y(x)dx}, \quad
    L_{{\frac{1}{2}}}^{cl}=\frac{1}{4}.
\end{equation}
So, we deduce the convergence of the first invariant estimate.\\

For $\gamma=\dfrac{3}{2}$ we have  an analog of the Plancherel
identity for the Fourier transform:
\begin{equation}
    \lim_{\chi \rightarrow +\infty}{\frac{S_{\frac{3}{2},0}(\chi
    y)}{\chi^2}}= L_{{\frac{3}{2}}}^{cl}
    \int_{\mathbb{R}}{y(x)^2dx}, \quad
    L_{{\frac{3}{2}}}^{cl}=\frac{3}{16}.
\end{equation}
Therefore, we get the convergence of the second invariant.

\section{Application of the SBSA to the ABP waves}

\subsection{ABP reconstruction}
In the previous subsection, we presented a new signal analysis
method based on the scattering transform. Now, we propose to use
this method for ABP analysis. For convenience we
replace the space variable $x$ by the time variable $t$.\\

We note the ABP signal $P(t)$ and the estimated pressure with the
SBSA technique $\hat{P}(t)$ such that:
\begin{equation}\label{pestime}
    \hat{P}(t)=\frac{4}{\chi}\sum_{n=1}^{N_\chi}{\kappa_{\chi,n}\psi_{\chi,n}^2(t)},
\end{equation}
where $-\kappa_{\chi,n}^2$ and $\psi_{\chi, n}$,
$n=1,\cdots,N_\chi$ are the $N_\chi$ negative eigenvalues and
associated $L^2-$normalized eigenfunctions of $H(-\chi P)$. We
recall that each component $4\kappa_{\chi,n}\psi_{\chi,n}^2$ in
(\ref{pestime}) refers to a single soliton \cite{GaGrKrMi:74},
\cite{LaCrSo:07}.\\

\begin{figure}[tb]
 \begin{center}
 \includegraphics[width=8cm]{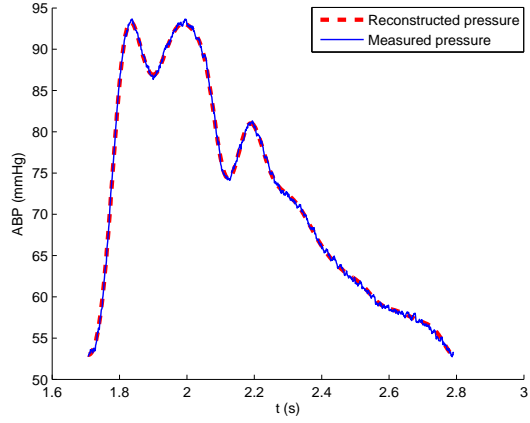}
  \caption{Measured and reconstructed pressure at the aorta}\label{pressions_solitons-aorte}
\end{center}
\end{figure}

\begin{figure}[tb]
 \begin{center}
 \includegraphics[width=8cm]{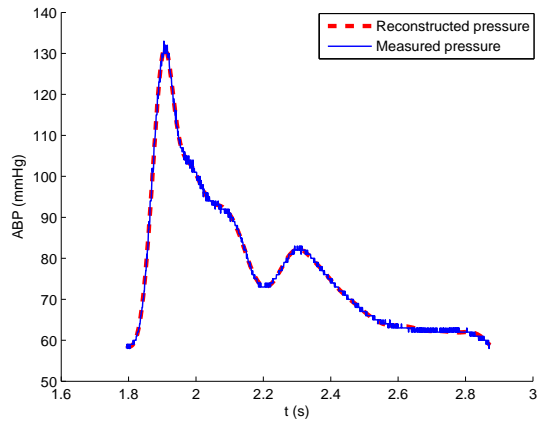}
  \caption{Measured and reconstructed pressure at the finger}\label{pressions_solitons-doigt}
\end{center}
\end{figure}

In Fig.~\ref{pressions_solitons-aorte} and
Fig.~\ref{pressions_solitons-doigt}, measured and reconstructed
pressures at the aorta and at the finger levels  are presented
respectively. The aortic pressure was measured using a catheter
while a Finapres was used to measure the pressure at finger. Only
5 to 10 components are sufficient for a good
reconstruction of the ABP waveform.\\

\subsection{Separation of the systolic and diastolic phases}
Following the work done in \cite{CrSo:07}, \cite{LaCrSo:07J}, we
propose here using the SBSA technique to separate the pressure
into fast and slow parts corresponding respectively to the
systolic and diastolic phases. Indeed a reduced model of ABP has
been proposed in \cite{CrSo:07}, \cite{LaCrSo:07J}. The latter
consists of a sum of two terms: a 2 or 3-soliton solution of a KdV
equation describing fast phenomena which predominate during the
systolic phase and a 2-element windkessel model describing slow
phenomena during the diastolic phase. As noticed in the previous
section, the SBSA technique decomposes the ABP signal into a sum
of solitons, each one characterized by its velocity given by the
discrete eigenvalues $-\kappa_{\chi,n}^2$. So the largest
$\kappa_{\chi,n}^2$, $n=1,\cdots,N_{s}$ describe fast phenomena
while the smallest ones describe slow phenomena. Referring to
\cite{CrSo:07}, \cite{LaCrSo:07J}, we take $N_{s}=2$ or $3$. We
note $\hat{P}_s$ and $\hat{P}_d$ the estimated systolic and
diastolic pressures respectively such that:
\begin{equation}
    \hat{P}_s(t)=\frac{4}{\chi}\sum_{n=1}^{N_{s}}{\kappa_{\chi,n}\psi_{\chi,n}^2},\quad
    \hat{P}_d(t)=\frac{4}{\chi}\sum_{n=N_{s}+1}^{N_\chi}{\kappa_{\chi,n}\psi_{\chi,n}^2}.
\end{equation}
We compute the first two invariants of these partial pressures
with the Riesz means for the chosen cut-off speed $\lambda$:
\begin{equation}
INV_1(\lambda)=\frac{4}{\chi}
S_{\frac{1}{2},\lambda}(\chi P),
\quad
INV_2(\lambda)=\frac{16}{3\chi^2}S_{\frac{3}{2},\lambda}(\chi P),
\end{equation}
We can now define the proposed invariants for the whole beat and
for the systolic and diastolic phases ($INV_j$, $INVS_j$,
$INVD_j$, $j=1,2$ respectively):
\begin{eqnarray}
\left\{%
\begin{array}{lc}
  INV_j = INV_j(0), &  \\
  INVS_j=INV_j(\lambda_s), &\quad j=1,2, \\
  INVD_j=INV_j(0)-INVS_j(\lambda_s), &  \\
\end{array}%
\right.
\end{eqnarray}
where $\lambda_s=\lambda_{\chi,2}$ or $\lambda_{\chi,3}$.\\

\begin{figure}[tb]
 \begin{center}
 \includegraphics[width=8cm]{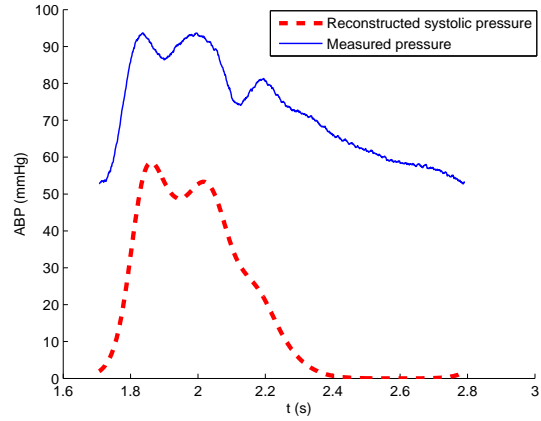}
  \caption{$\hat{P}_s$ and fast systolic phenomena}\label{soliton_part}
\end{center}
\end{figure}

\begin{figure}[tb]
  \begin{center}
\includegraphics[width=8cm]{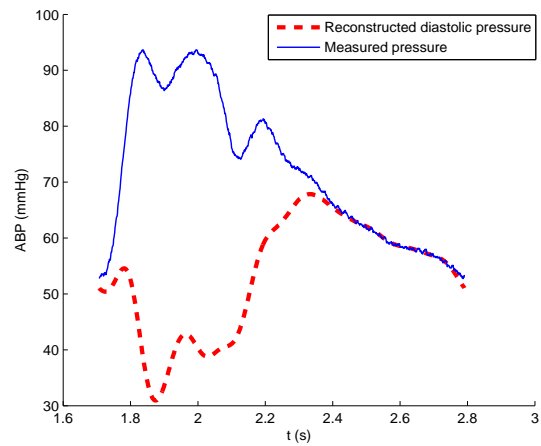}
  \caption{$\hat{P}_d$ and slow diastolic phenomena}\label{WK_part}
\end{center}
\end{figure}

In Fig.~\ref{soliton_part} and Fig.~\ref{WK_part}, we represent
the measured pressure and the estimated systolic and diastolic
parts respectively. We remark that $\hat{P}_s$ and $\hat{P}_d$ are
respectively localized during the systole and the diastole, as
expected.\\

\section{New spectral cardiovascular indices}
\subsection{The SBSA parameters and the ABP}
The SBSA technique provides a new description of the ABP signal
using the DST. As seen in the previous section, the reconstruction
of the signal by IST from its spectral data gives good results.
Instead of reconstructing the original signal, it is possible to
modify the spectrum leading to some kind of filtering. This is
illustrated by the separation of the systolic and diastolic
phases. Moreover, the SBSA method introduces new parameters. The
first two global invariants $INV_1$ and $INV_2$ are respectively,
by definition, the usual mean blood pressure (MBP) and the less
usual, but easy to compute directly, integral of the square of the
pressure. The first systolic invariant $INVS_1$ is a new index: we
think that it can be correlated to SV. Remark that if $INV_2$ is
easy to compute directly, the "fast part" of this integral, the
second systolic invariant $INVS_2$ is a new less obvious index and
might contain information on ventricular contractility. SV and
contractility are in fact parameters of great interest that are
difficult to measure routinely, as they require invasive or
sophisticated techniques. For instance SV can be estimated by
invasive nuclear ventriculography \cite{Wi:89}, 2D
echocardiography \cite{Ko:03}, radionuclide monitoring
\cite{Ta:01}, impedance cardiography \cite{Ch:00}, \cite{Cu:97}.
Only one evaluation of SV  from ABP has been proposed
\cite{Ha:99}. Ventricular contractility is assessed by the mean of
the tissue doppler echocardiography \cite{Go:97}, \cite{Gr:02}.\\

\begin{figure}[tb]
 \begin{center}
  \includegraphics[width=8cm]{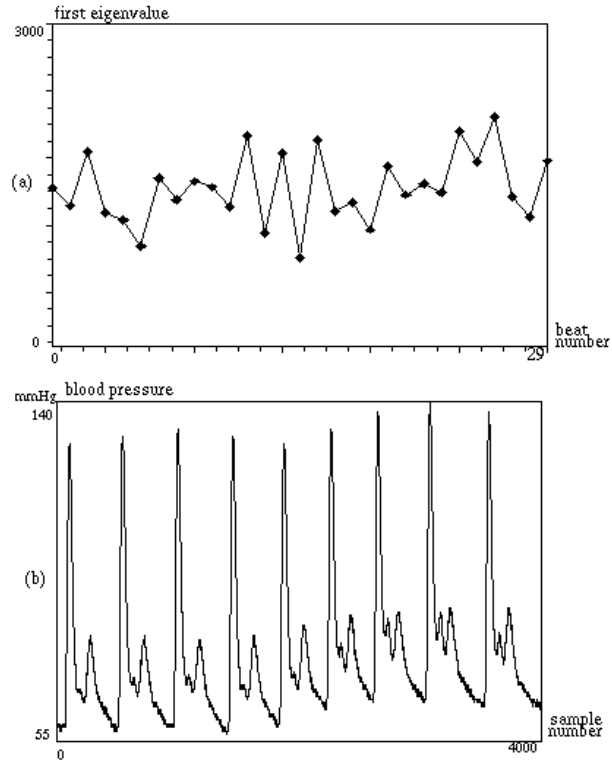}
  \caption{(a) First eigenvalue time series of a healthy subjects under $0.25Hz$ paced-breathing during 29 beats.
  The signal exhibits a beat-to-beat variability that is twice as fast as the breathing rate.
  (b) A zoom on a few beats shows morphological changes in the blood pressure signal.}\label{VHF}
\end{center}
\end{figure}

On the other hand, the eigenvalues computed with the SBSA
technique are strongly dependent upon the ABP waveform. Indeed,
when the pressure waveform changes, the optimal value of the
parameter $\chi$ and the eigenvalues also change. This fact is
illustrated in Fig. \ref{VHF}. Therefore, the eigenvalues could be
used to assess the baroreflex sensitivity (BRS) in a certain way.
In fact, the BRS expresses the variation of the heart beat
interval in response to each arterial pressure variation. The BRS
concept was first based on drug-induced responses of ABP and heart
period \cite{Pa:92}. Then various time \cite{Li:03}, \cite{Mo:02},
\cite{Pa:95} and spectral \cite{Bo:04}, \cite{MaMoMe:01},
\cite{Mo:02}, domain methods \cite{Li:03} were compared.\\

In this section we analyse the SBSA parameters in two situations,
devoted to the assessment of the ANS control of the CV system. We
restrict the study to the modulus of the  first two negative
eigenvalues $|\lambda_{\chi,1}|=|\lambda_{1}|$,
$|\lambda_{\chi,2}|=|\lambda_{2}|$ and the first two invariants
(global, systolic and diastolic). We include in the analysis some
classical parameters which are PI, SBP and DBP.\\

\subsection{Head-up 60 degrees tilt-test}
The head-up tilt test is mainly used for vasovagal syncopes
diagnosis, characterized by an autonomic dysfunction. It consists
in the orthostatic transition from the supine to the standing
positions. This leads to a redistribution of the venous blood
volume, from the intrathoracic region towards the venous volume in
the leg and lower abdominal veins. This leads to a decrease in SV
and PI and an increase in SBP \cite{Cu:97},
\cite{Sh:99}, \cite{Si:04}, \cite{Ta:07}.\\

A group of 15 healthy subjects under $0.25 Hz$ paced breathing,
already studied \cite{Be:98}, was considered. The table was
rotated to an upright position at 60 degrees.  The continuous ABP
was measured at the finger using a Finapres device \cite{We:85}.
The two positions, supine and standing, were compared using the
Wilcoxon non parametric paired test.\\

\begin{figure}[tb]
\begin{center}
\includegraphics[width=10cm]{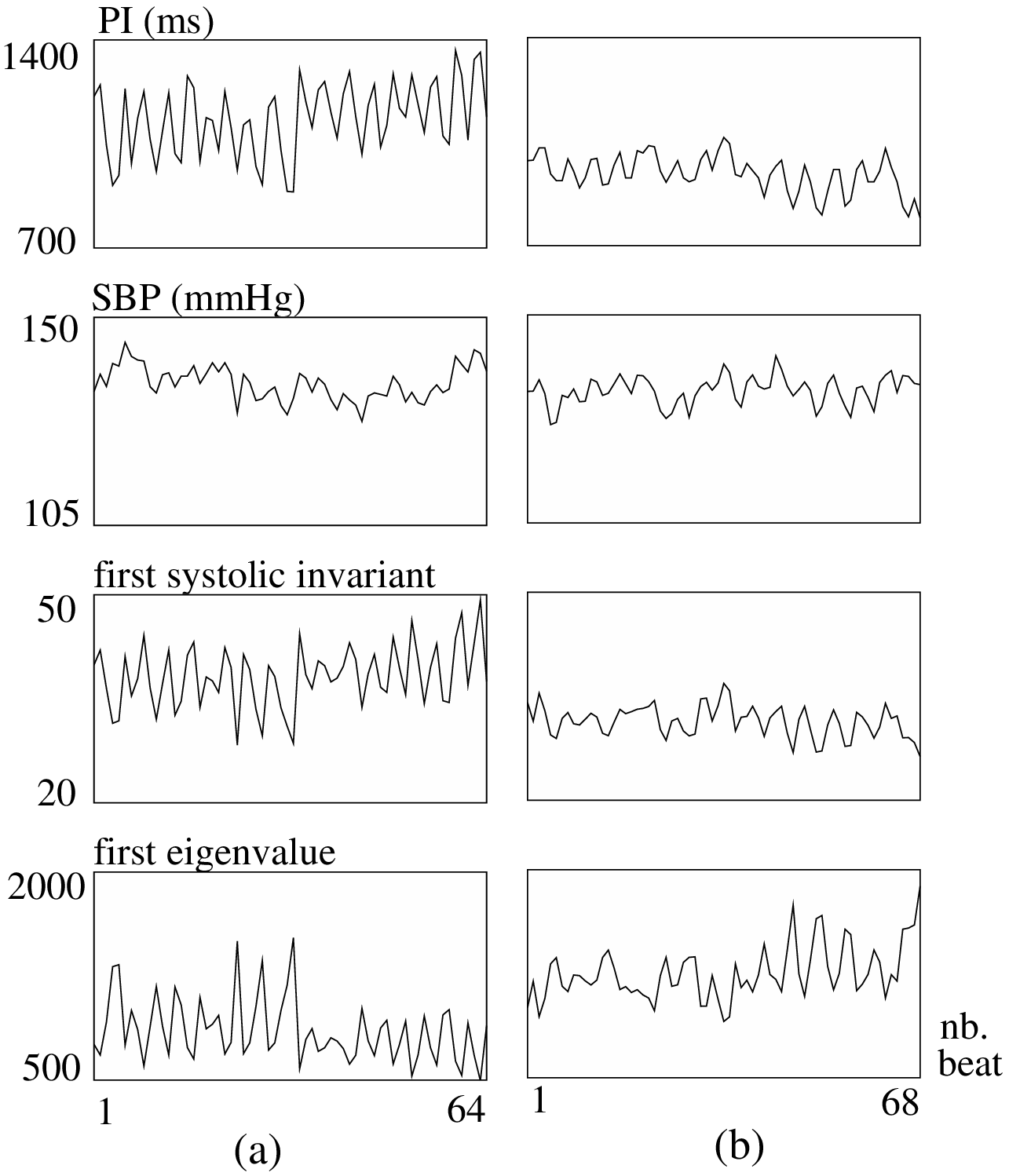}
\caption{Time series of ABP parameters in a healthy subject under
$0.25 Hz$ paced breathing, in  supine (a) and standing (b)
positions after a 60 degrees  tilt test. PI and $INVS_1$ are
reduced whereas $|\lambda_1|$ is increased in the standing
position, more than two minutes after the tilt test. SBP has the
same level in both positions.} \label{signaux-tilt}
\end{center}
\end{figure}

Fig. \ref{signaux-tilt} shows the time series of PI, SBP, $INVS_1$
and $|\lambda_1|$ in the supine and standing positions. Mean
levels of the ABP parameters are presented in Table
\ref{table:mean-tilt2}. We notice that significant differences
between the supine and standing positions appear for PI, DBP,
$|\lambda_1|$, $|\lambda_2|$ and $INVS_1$ while for the other
parameters the differences are not significant. So after the tilt
test, SBP and MBP recover their prior values (preceding the tilt
phase) whereas $INVS_1$ is decreased. If $INVS_1$ is related to SV
then this means that SV remains decreased after the tilt test and
this can be explained by the decrease in the venous return.\\


\begin{table}[tb]
 \begin{center}
  \caption{ ABP parameters during tilt protocol in 15 healthy subjects}\label{table:mean-tilt2}
\footnotesize
\begin{tabular}{lcccc}
  \hline
    ABP parameters & &supine& standing&probability \\
 \hline
   Direct&\multicolumn{4}{c}{  } \\
  \hspace{0.5cm} PI  && $ 918 \pm 33 $ & $ 778 \pm 21$& ***\\
   \hspace{0.5cm}SBP && $ 121 \pm 4  $ & $ 127 \pm  3$& NS\\
   \hspace{0.5cm}DBP && $ 68  \pm 4  $ & $ 80  \pm  3$& **\\
\hline
  Eigenvalues&\multicolumn{4}{c}{  }\\

   \hspace{0.5cm}First   && $ 1515 \pm 131 $ & $ 1965 \pm137 $&***\\
   \hspace{0.5cm}Second  && $ 1205 \pm 86  $ & $ 1612 \pm 92 $&***\\
\hline
  First  invariants&\multicolumn{4}{c}{  }\\

   \hspace{0.5cm}Global    && $ 77 \pm 6  $&$ 71 \pm 4 $& NS\\
   \hspace{0.5cm}Systolic  && $ 21 \pm 1  $&$ 19 \pm 1 $&**\\
   \hspace{0.5cm}Diastolic && $ 56 \pm 4  $&$ 51 \pm 3 $&NS\\
\hline
   Second invariants&\multicolumn{4}{c}{  }\\
   \hspace{0.5cm}Global    && $ 7041 \pm 975 $&$ 6833 \pm 711$&NS\\
   \hspace{0.5cm}Systolic  && $ 2705 \pm 328 $&$ 2566 \pm 247$&NS\\
   \hspace{0.5cm}Diastolic && $ 4336 \pm 648 $&$ 4276 \pm 465$&NS\\
\hline \multicolumn{5}{l}{Data are expressed as means and SEM; **:
$P\leq .01$; ***: $P\leq .001$}
\end{tabular}\end{center}
\end{table}

\begin{figure}[tb]
\begin{center}
\includegraphics[width=10cm]{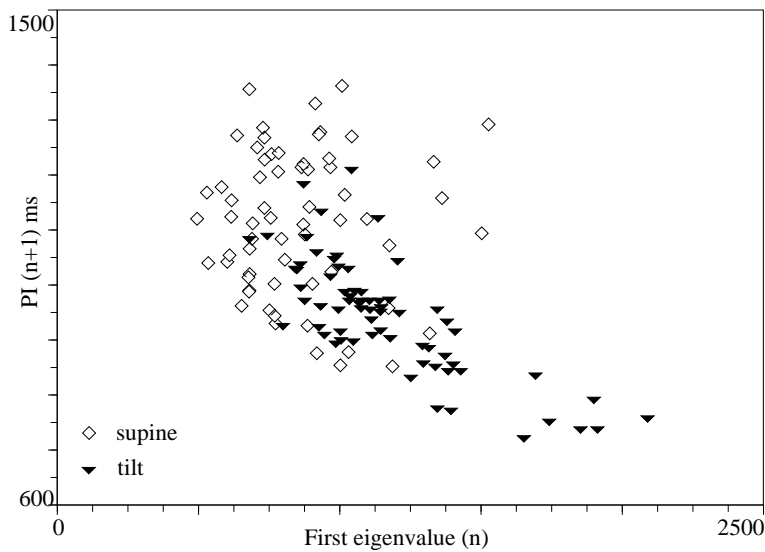}
\caption{Beat-to-beat BRS, represented as the relation between
$|\lambda_1(n)|$ and $PI(n+1)$. Slope and correlation are stronger
during tilt.  } \label{baro-tilt}
\end{center}
\end{figure}

The ABP analysis during the transition from the supine to the
standing positions was possible for only nine subjects. For these,
we assessed the linear  relation between PI of the beat $n+1$ and
$|\lambda_1|$ computed for the beat $n$. In Fig. \ref{baro-tilt},
we note that as PI decreases, $|\lambda_1|$ increases and that the
linear correlation   between PI and $|\lambda_1|$ is stronger
around the transition than in supine position. Table
\ref{table:baro-tilt} shows that the correlation coefficient
($R^2$) and the slope, compared by a one-way repeated measures
analysis of variance, were significantly stronger around the
transition than in supine or standing positions. Moreover, a
comparison with usual BRS indices, SBP and pulse pressure (PP),
shows that $|\lambda_1|$ has the strongest correlation with PI
(Table \ref{table:baro-3estimators-tilt}). It is not surprising
that $|\lambda_1|$ is more informative about the relation between
the heart period and the ABP because it reflects the arterial
waveform and not only one (SBP) or two (PP) samples of this
waveform.\\

\begin{table}[tb]
\footnotesize \caption{Beat-to-beat BRS during  tilt protocol }
\label{table:baro-tilt}
\begin{center}
\begin{tabular}{|c|c|c|c|c|}\hline
&supine&tilt&standing& probability\\
\hline
$R^2$&$.335 \pm .102$& $.611 \pm .074$&$.421 \pm .079$& ***\\
\hline
slope&$-.105 \pm .03$&$-.177 \pm .03$&$-.106\pm.015$& **\\
\hline
\end{tabular}
\end{center}
\footnotesize {$R^2$ and slope of the linear regression between
$|\lambda_1(n)|$ and $PI(n+1)$, over about 60 beats, in 9 healthy
subjects. Data are expressed as means and SEM; ** $p\leq .01$; ***
$p\leq .001$, at repeated measures analysis of variance. Slope and
$R^2$ are the strongest during tilt.}
\end{table}

\begin{table}[tb]
\footnotesize \caption{Three indices of beat-to-beat BRS during
tilt } \label{table:baro-3estimators-tilt}
\begin{center}
\begin{tabular}{|c|c|}\hline
$R^2$&tilt\\
\hline
$|\lambda_1|$&$.611 \pm .074$\\
\hline
SBP&$.351 \pm .091$\\
\hline
PP&$.415 \pm .069$\\
\hline
\end{tabular}
\end{center}
\footnotesize {Mean of  $R^2$  of the linear regression between
ABP parameters ($|\lambda_1|$, SBP, PP) and $PI$, over about 60
beats, in 9 healthy subjects of the tilt test. Data are expressed
as means and SEM. $|\lambda_1|$ is the most strongly correlated
with PI.}
\end{table}

\subsection{Isometric handgrip exercise}
The isometric handgrip exercise is mainly used for the evaluation
of a non appropriate ANS behavior, that mimics actual conditions
of professional or domestic exposure, with arterial hypertension.
Indeed, the ANS acts in the same way as in a dynamic exercise
usually characterized by an increase in muscle oxygen needs. The
isometric exercise is a form of exercise involving the static
contraction of a muscle without any visible movement in the
angle of the joints \cite{Fr:70}.\\

A group of 13 healthy subjects was considered. The continuous ABP
was measured at the finger with a Finapres device \cite{We:85}.
The two conditions: at rest and during the handgrip test, were
compared using the Wilcoxon  non parametric paired test.\\

\begin{figure}[tb]
\begin{center}
\includegraphics[width=10cm]{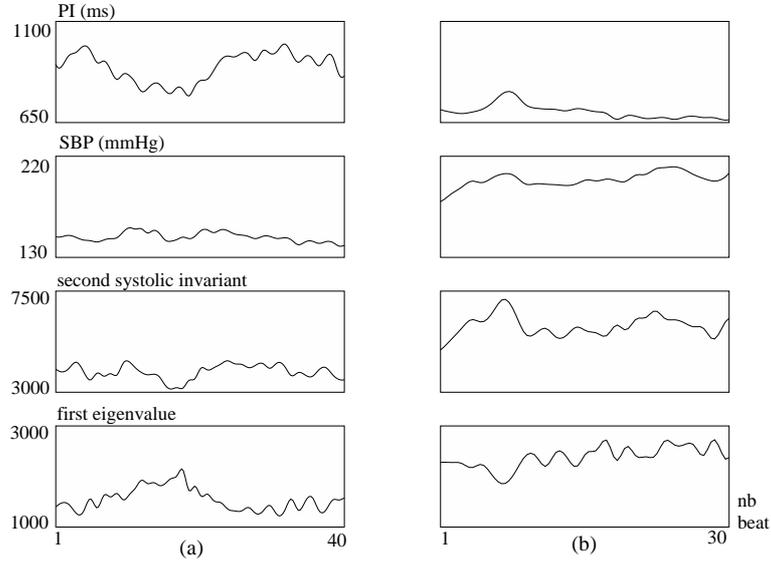}
\caption{Time series of ABP parameters in a healthy subject under
spontaneous breathing, at rest (a) and  during the  handgrip test
(b). PI is reduced whereas SBP, $INVS_2$ and $|\lambda_1|$ are
strongly increased during handgrip.} \label{signaux-hg}
\end{center}
\end{figure}

In Fig. \ref{signaux-hg}, the time series of PI, SBP, $INVS_2$ and
$|\lambda_1|$ are presented at rest and during the handgrip test.
Table \ref{table:mean-hg} illustrates the mean levels of the ABP
parameters. We notice that while all the first invariants do not
change significantly, the second invariants are sensitively
increased during handgrip.\\

The voluntary central command, involved in the handgrip,
synchronously activates the motor and CV systems, leading first to
an increase in the heart rate, followed by an increase in the ABP
\cite{Ho:82}, \cite{McClo:81}. So, as expected, PI decreases and
SBP increases (Table \ref{table:mean-hg}). Moreover, $INVS_1$
might inform us that SV does not change while $INVS_2$ might
inform us about the increasing contractility, as if the heart
tries to eject the same quantity of blood  during a smaller
period. Such a result evokes the \emph{treppe effect} (or
frequency-force
relation), where an increase in heart rate indirectly induces an increase in contractility.\\

\begin{table}[tb]
 \begin{center}
\footnotesize{
  \caption{ABP parameters during handgrip protocol in 13 healthy
subjects}\label{table:mean-hg}
\begin{tabular}{lcccc}
  \hline
  ABP parameters & & rest& handgrip &probability \\
 \hline
   Direct&\multicolumn{4}{c}{  } \\
  \hspace{0.5cm}PI&   & $ 953\pm 59$&$748\pm 42$&***\\
   \hspace{0.5cm}SBP&& $ 148\pm 5$&$183\pm 5$&***\\
   \hspace{0.5cm}DBP &&$ 83\pm 3 $&$104\pm 4 $&*** \\
\hline
  Eigenvalues&\multicolumn{4}{c}{  }\\

   \hspace{0.5cm}First &  &$1439\pm 160  $&$2325 \pm 249$&**\\
   \hspace{0.5cm}Second & &$1186\pm 125  $&$1872 \pm 180 $&***\\
\hline
  First  invariants&\multicolumn{4}{c}{  }\\

   \hspace{0.5cm}Global &   & $96\pm6 $&$ 95\pm 6 $  &NS\\
   \hspace{0.5cm}Systolic&  & $26\pm 1 $&$26\pm 1 $  &NS\\
   \hspace{0.5cm}Diastolic& & $70\pm 4 $&$69\pm 4 $  &NS\\
\hline
   Second invariants&\multicolumn{4}{c}{  }\\
   \hspace{0.5cm}Global  &  & $10239\pm866 $&$12886 \pm1192 $&**\\
   \hspace{0.5cm}Systolic & & $3956\pm344 $&$5013 \pm 470$  &**\\
   \hspace{0.5cm}Diastolic& & $6283\pm526 $&$7872 \pm727 $&**\\
\hline \multicolumn{5}{c}{Data are expressed as means and SEM; **
$p\leq .01$; *** $p\leq .001$. }
\end{tabular}}
\end{center}
\end{table}

\begin{figure}[tb]
\begin{center}
\includegraphics[width=10cm]{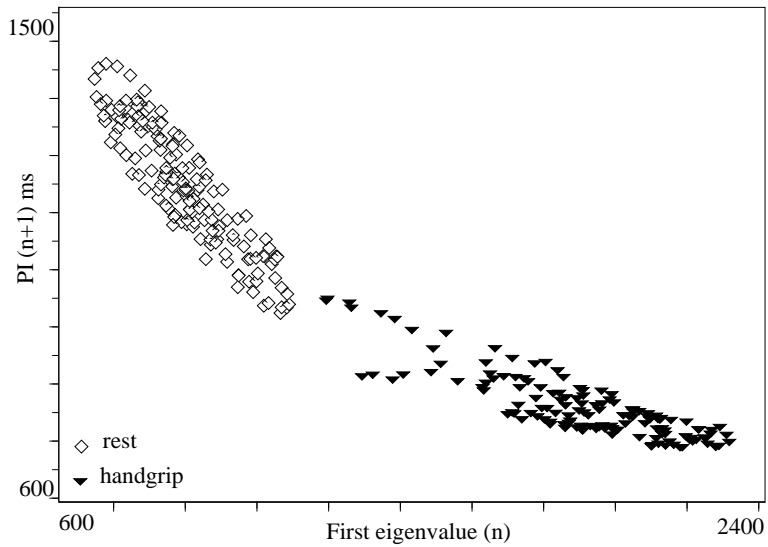}
\caption{Beat-to-beat BRS, represented as the relation between
$|\lambda_1(n)|$ and $PI(n+1)$. Slope is lower during handgrip. }
\label{baro-hg}
\end{center}
\end{figure}

As in the case of the tilt test, we study  the relationship
between PI and $|\lambda_1|$. Fig. \ref{baro-hg} illustrates the
relation between PI of the beat $n+1$ and $|\lambda_1|$ computed
for the beat $n$ at rest and during the handgrip test. The  strong
linear correlation between PI and $|\lambda_1|$ is the same at
rest and during the handgrip,  but the slope is significantly
lower during the handgrip (Table \ref{table:baro-hg}). Moreover, a
comparison with usual  BRS indices, SBP and PP, shows that
$|\lambda_1|$ has the strongest correlation with PI (Table
\ref{table:baro-3estimators}). This result, as obvious as in the
case of the tilt test leads us to consider $|\lambda_1|$ as a
promising index that can be used to study the relation between the
heart period and ABP.\\

\begin{table}[tb]
\footnotesize \caption{Beat-to-beat BRS during  handgrip protocol
} \label{table:baro-hg}
\begin{center}
\begin{tabular}{|c|c|c|c|}\hline
&rest&handgrip& probability\\
\hline
$R^2$&$.647 \pm .054$&$ .609 \pm .050$& NS\\
\hline
slope&$-.302 \pm .071$&$-.132 \pm .024$& **\\
\hline
\end{tabular}
\end{center}
\footnotesize { $R^2$ and slope of the linear regression between
$|\lambda_1(n)|$ and $PI(n+1)$, over about 40 beats, in 13 healthy
subjects. Data are expressed as means and SEM; ** $p\leq .01$; ***
$p\leq .001$, at  paired test. Slope is reduced during handgrip
whereas  $R^2$ is not significantly different. }
\end{table}

\begin{table}[tb]
\footnotesize \caption{Three indices of beat-to-beat BRS during
handgrip protocol} \label{table:baro-3estimators}
\begin{center}
\begin{tabular}{|c|c|c|}\hline
$R^2$&rest&handgrip\\
\hline
$|\lambda_1|$&$.647 \pm .054$&$ .609 \pm .050$\\
\hline
SBP&$.166 \pm .055$&$.261\pm .060$\\
\hline
PP&$.375 \pm .078$&$.284\pm .060$\\
\hline
\end{tabular}
\end{center}
\footnotesize { Mean $R^2$  of the linear regression between ABP
parameters ($|\lambda_1|$, SBP, PP) and $PI$, over about 40 beats,
in the 13 healthy subjects of the handgrip test. Data are
expressed as means and SEM. $|\lambda_1|$ is the most strongly
correlated with PI.  }
\end{table}

\section{Conclusion}
This article deals with a new ABP analysis method based on the
scattering theory.  This SBSA method can be thought of as a
nonlinear Fourier analysis for pulse-like signals. It allows
analysis and precise reconstruction as is shown by the very good
agreement between real and estimated pressures. We have also
presented an application to a filtering problem consisting in
separating the systolic and diastolic phases. Then, we have
introduced new cardiovascular indices computed with the SBSA
method. These parameters include the first two systolic invariants
and  we think that they might give information on the variation of
the stroke volume and the ventricular contractility, that are
difficult to measure routinely. Another interesting parameter is
the first eigenvalue which seems to reflect the BRS in a certain
way. The results obtained from the analysis of two widely used
physiological conditions are promising and we are now working
on the validation of the advanced hypotheses.\\

\newpage
\bibliography{references}
\end{document}